\documentclass[aps,floatfix,onecolumn]{revtex4-2}

\usepackage{amsmath,amssymb}
\usepackage{graphicx}
\usepackage{epsfig}
\usepackage[usenames]{color}
\usepackage{subcaption}
\usepackage{hyperref}
\usepackage{natbib}

\usepackage{xcolor}
\usepackage{graphicx}%   Include figure files
\usepackage{gensymb}
\usepackage{dcolumn}%   Align table columns on decimal point
\usepackage{bm}%   bold math

\makeatletter
\newcommand*{\addFileDependency}[1]{%   argument=file name and extension
  \typeout{(#1)}
  \@addtofilelist{#1}
  \IfFileExists{#1}{}{\typeout{No file #1.}}
  }
\makeatother

\newcommand*{\myexternaldocument}[1]{%  
  \externaldocument{#1}%  
  \addFileDependency{#1.tex}%  
  \addFileDependency{#1.aux}%  
  }
\usepackage{xr}
\myexternaldocument{supplementary}
\usepackage{xspace}

\begin{document}

\title{Approach to Data Science 
%and Generalized Quantum Dynamics of Particles  with ٪the Use of 
with Multiscale  Information Theory  }%Generalized Nonlinear Schr\"{o}dinger Equation
\author{Shahid Nawaz \thanks{ snafridi@gmail.com}}
\affiliation{\small 18 Lake Shore Dr. Apt 2B, Watervliet, NY, 12189, USA.}
%}
\
%\author[1]{Shahid Nawaz \thanks{ snafridi@gmail.com}}
%\author[2]{Muhammad Asjad\thanks{muhammad.asjad@ku.ac.ae}}
%Department of Physics, Bellarmine University\\
 %2001 Newburg Road, Louisville, KY 40205, USA\\
%email: \url{msaleem@bellarmine.edu}
\author{Muhammad Saleem\thanks{msaleem@bellarmine.edu}}
\affiliation{\small Department of Physics, Bellarmine University, 2001 Newburg Road, Louisville, KY 40205, USA }

\author{F. V. Kusmartsev %\thanks{Corresponding Author: dalaver.anjum@ku.ac.ae}
}
\affiliation{\small Department of Physics, Khalifa University, PO Box 127788, Abu Dhabi, United Arab Emirates}
\author{Dalaver H.~Anjum\thanks{Corresponding Author: dalaver.anjum@ku.ac.ae}}
\affiliation{\small Department of Physics, Khalifa University, PO Box 127788, Abu Dhabi, United Arab Emirates}

%\affil[1]{\small 18 Lake Shore Dr. Apt 2B, Watervliet, NY, 12189, USA.}
%\affil[2]{Address }
%\affil[3]{address}
%EndAName
%{\small 118 Yardboro Avenue, Albany, NY, USA. }\\

\begin{abstract}\label{abs}

Data Science is a multidisciplinary field that plays a crucial role in extracting valuable insights and knowledge from large and intricate datasets. Within the realm of Data Science, two fundamental components are Information Theory (IT) and Statistical Mechanics (SM), which provide a theoretical framework for understanding dataset properties. IT enables efficient storage and transmission of information, while SM focuses on the behavior of systems comprising numerous interacting components. In the context of data science, SM allows us to model complex interactions among variables within a dataset. By leveraging these tools, data scientists can gain a profound understanding of data properties, leading to the development of advanced models and algorithms for analysis and interpretation. Consequently, data science has the potential to drive accurate predictions and enhance decision-making across various domains, including finance, marketing, healthcare, and scientific research.

In this paper, we apply this data science framework to a large and intricate quantum mechanical system composed of particles. Our research demonstrates that the dynamic and probabilistic nature of such systems can be effectively addressed using a Multiscale Entropic Dynamics (MED) approach, derived from the Boltzmann methods of SM. Through the MED approach, we can describe the system's dynamics by formulating a general form of the Nonlinear Schr\"{o}dinger equation and how it can be applied to various systems with particles and quasi-particles, such as electrons, plasmons, polarons, and solitons. By employing this innovative approach, we pave the way for a deeper understanding of quantum mechanical systems and their behaviors within complex materials.

\end{abstract}
\maketitle
\section{Introduction}\label{intro}

The fields of statistical physics, information theory, and data science have many commons and they are interconnected in several ways\cite{jaynes1957information}. These three fields deal with the analysis of complex systems, and they use similar mathematical and statistical tools to model and analyze these systems.

One of the important connections between these fields is the concept of entropy \cite{wehrl1978general}. In statistical physics, entropy is a measure of the disorder or randomness of a physical system and can be used as the main driving force of the Second Law of thermodynamics\cite{presse2013principles}. In information theory, entropy is a measure of the uncertainty or randomness of a message or information content of a data set. In data science, entropy is often used to measure the randomness or unpredictability of data sets or in classical and quantum models\cite{kaufmann2020discovery}.

Another connection between these fields is the use of probability theory and statistical methods to model and analyze complex systems\cite{caticha2008lectures}. Statistical physics uses probability theory to model the behavior of large systems of particles, while information theory
 uses probability theory to quantify the uncertainty or randomness as well as a measure of information of messages or data sets\cite{mezard2009information}. Data science uses statistical methods to model and analyze data sets and to develop algorithms for processing and interpreting data\cite{tan2020granularity}.

{ On the other hand, quantum mechanics have been conventionally formulated in the Hilbert Space with the use of two conjugate variables that obey Heisenberg's uncertainty principle. For instance, the momentum $p$ and the position coordinate $q$, in the form of $\Delta q \Delta p \geq\hbar/2$. In this way, this principle describes the statistical nature of these self-adjoint operators $\hat{q}$ and $\hat{p}$ in Hilbert Space. Whereas in the entropic dynamics (ED) formulation, the quantum nature of these operators is given a secondary role. In fact, in this approach, the uncertainty in these variables stems from the diffusion process of the Brownian motion of the particles \cite{nawaz2012momentum}.}

{
It is important to note that in ED momentum is not real, it is not ontic. 
Momentum is an epistemic property of wave functions and not a property of the particles.
In classical mechanics, we assume that the particles' positions and their momenta are real. In ED only the positions are real. 
The reason for this is that ED is not a dynamic of particles. it is a dynamic of probabilities. 
This is important: in this dynamics, there is a momentum that is canonically conjugate to the generalized coordinates,
 but the latter are the probabilities and not the positions. Thus, there is no moment for the particles. 
Of course, we can always consider translations, and, following convention, 
we can call the generator of translations by the name "linear momentum".
 But this is just a name for operators which are not properties of the particles. 
}

%\textcolor{red}{ FK- was not sure the meaning of this phrase:}
%\textcolor{blue}{ We explain it again in the above}

The ED approach is an alternative formulation of QM that puts on the first hand the probabilistic nature of all physical quantities %FK-do not understand it: and first, and the dynamics second 
\cite {Caticha_2011, e21100943}. This approach seems to describe %resolve the discord between 
the dynamics and probabilistic nature of quantities in a more convenient way than other approaches. 
The ED formulation of quantum mechanics was introduced in 2009 and has been applied to fields such as quantum measurement problem \cite{johnson2012entropic, vanslette2017quantum}, uncertainty relations \cite{nawaz2012momentum} curved space-time \cite{nawaz2016entropic}, scalar fields \cite{ipek2015entropic, Ipek_2019, ipek2020entropic}, and finance \cite{bai2020entropic}.

The paper below is organized as follows. In section~\ref{ED-form-QM} the ED formulation of QM is briefly presented. The approach to data science using MED is discussed in section~\ref{DS-MED-method}.  In section~\ref{ED-sec1} %the GSE is developed under
we illustrate how the MED framework can be used to derive  GSE. In section \ref{solitons-1}, the derived GSE is adopted %reduced 
to a couple of well-known forms that describe the dynamics of nonlinear phenomena. %non-interacting and interacting
%. 
Furthermore, a form for the interaction of electromagnetic fields with solitons or polarons is also presented. Here we assumed that solitons like polarons have self-trapped particles  and therefore  they are charged, of course. The impact and ramifications of the GSE in the context of describing the dynamics of solitons in 2d materials are presented in the  section \ref{Discussion}. In the end, the overall conclusions drawn from the presented examples of the application of MED method to various complex systems are presented and discussed in the context of Data Science, see the section \ref{conclusions}.

\section{ ED formulation of QM}\label{ED-form-QM}

%The ultimate goal is to derive quantum mechanics (QM) as an application of ED.
One notes that the fundamental object in QM is the wave function $\Psi$, which is a complex function of coordinate or momenta. It may, generally, be represented  with the use of  %that further involves 
two real quantities $\rho$ and $\Phi$: $\Psi=\sqrt{\rho} \exp{(i \Phi)}$. Here $\rho$ is the probability density and $\Phi$ is the phase field. 
The wave function is well described by $U(1)$ gauge symmetry group, where the phase $\Phi$ is a free gauge parameter. The elements of the abelian gauge group $U(1)$ are associated with the different values of the phase. 

It turns out that these two quantities are the degrees of freedom in ED, which are determined via two sectors.
%% FK: degrees of freedom are usually variables, it does not look to me as variables so far???
The first sector involves the entropy functional \eqref{entropy-fun} that leads to the Fokker-Planck (FP) equation \eqref{FP1}. The second sector involves an energy functional \eqref{H1}, whose conservation leads to quantum Hamilton-Jacobi equation \eqref{HJ2}. The two equations when combined give the Schr\"{o}dinger equation.

%The transition probability given in equation (7) represents a general form that will be reduced to a Gaussian form under the formulation of Brownian motion (reference..). It does not have time evolution but it is a prerequisite for finding the time evolution Gaussian form of probability density. Finding the time evolution form of general transition probability requires knowing a quantity called entropic time.  

It is pertinent to mention herein that in the past the Schr\"{o}dinger equation has been derived in  
\cite{nelson2021quantum} with the use of stochastic mechanics. The starting point in this derivation is the transition probability,  which is conventionally described by  a Greens function.  In that stochastic  formulation \cite{nelson2021quantum}  it is given below

%The particles were assumed to be undergoing the Brownian motion and as a consequence, the probability distribution in true Gaussian form was simply presented without any explicit derivation. Their calculations are given below.   

%The transition probability of Brownian motion in $d-$dimension is given by 
\begin{equation}
p(x^\prime, t^\prime|x, t)=\frac{1}{(2\pi \sigma^2\Delta t)^{d/2}}e^{-\frac{(x^\prime-x)^2}{2\sigma^2\Delta t}}\,,\label{wiener-process}
\end{equation}
where $d$ is the dimension of the space and $\sigma^2/2$ is the diffusion coefficient. The Brownian motion given by equation \eqref{wiener-process} keeps track of the future given the present is known while independent of its past. Equation \eqref{wiener-process} admits the following stochastic differential equation
\begin{equation}
\Delta x^i(t)=b^i(x(t),t)\Delta t+\Delta w^i(t)\,,\label{SDE}
\end{equation}
where $b^i$ is drift velocity and $w^i$ is the fluctuation or noise in the Brownian Motion with the following correlators
\begin{equation}
\langle\Delta w^i\rangle=0\,, \ \ \text{and}\ \ \langle\Delta w^i\Delta w^j\rangle=\sigma^2\Delta t\delta^{ij}\,,\label{fluctuations} 
\end{equation}

On the other hand, the transition probability (Greens function) is derived in the ED formulation followed by the Brownian motion. The ED framework differs from stochastic mechanics as follows.

{The ED formulation for a certain quantum system begins with defining the entropy functional \eqref{entropy-fun} subjected to relevant constraints of the system and defining the notion of entropic time. The relevant constraints are those that lead to the desired theory. Since our main concern is to derive quantum theory using ED, the relevant constraints are the phase constraint \eqref{phi-pot} and gauge constraint \eqref{vec-pot}. Once the constraints are incorporated and the entropy functional is optimized, one obtains the transition probability \eqref{Prob1}. Apparently, \eqref{Prob1} is timeless. But this is a common feature of Bayesian or entropic  inferences, where the goal is to update the prior probability to posterior probability when new information becomes available. The new information could be in the form of data (Bayesian inference) or in the form of constraints (entropic inference). In both cases, the two methods of inference are atemporal. It does not matter whether the posterior is obtained in the past or present, one gets the same result. Interestingly it is possible to introduce time in ED. Consider a particle that moves from the initial position $x$ to the final position $x^\prime$. Generally, both positions are unknown. This means that we are dealing with the joint probability $P(x, x^\prime)$. Then using the product rule of probability
\begin{equation}
P(x,x^\prime)=P(x^\prime|x)P(x)\,,\label{prod-rule}
\end{equation}
where $P(x^\prime|x)$ is the probability of $x^\prime$ given $x$. Since $x$ is also unknown, we marginalize over $x$
\begin{equation}
P(x^\prime)=\int dx P(x^\prime,x)=\int dx P(x^\prime|x)P(x)\,,
\end{equation}
where $P(x)$ is the probability of the particle being located at position $x$. Whereas $P(x^\prime)$ is the particle at position $x^\prime$. Since $x$ occurs at an initial instant $t$ and $x^\prime$ happens at a later instant $t^\prime$. Therefore we set 
\begin{equation}
P(x)=\rho(x,t)\ \ \text{and} \ \ P(x^\prime)=\rho(x^\prime, t^\prime)\,.
\end{equation}
In summary, in ED, time is introduced as a book-keeping device that keeps track of change. The notion of time would be further elaborated in section \ref{ED-sec1} where the duration of time would also be obtained.}

%to the system are expressed in the form of expected values or Lagrange Multipliers. Once the constraints are specified, they are incorporated into the entropy function prior to its optimization. The result of optimization represents a probability distribution. %It is pertinent to note that the determination of Lagrangian \cite{goldstein:mechanics} of the classical system has a similarity \textcolor{red}{????? what????} with the determination of the probability distribution of the quantum system. The former gives Newton's equation of motion for describing the dynamics of particles in a classical system. Whereas the latter gives the energy equation\textcolor{red}{FK: do not understand, that is equation primary  derived for quantum numbers, that what you mean, probabilistic was an interpretation of which many were confused} that addresses the dynamic and probabilistic nature of quantum mechanical quantities. \textcolor{red}{I prefer Heisenberg equation}
%\includegraphics[4cm]{Figure1.png}

\section{Approach to Data Science with  MED method}\label{DS-MED-method}

The emergence of new information technologies has led to the rise of data science, which involves the analysis of large and complex data sets to make predictions about the evolution of systems. Data science has become an essential tool for businesses and organizations to gain insights into their customers, products, and operations, and make data-driven decisions.

To conduct a data analysis, one must first take a data set and break it down into subsets. By analyzing the multiplicity of these subsets, one can determine the probabilities for the realization of different outcomes. This approach is based on the principles of probability theory, which involves quantifying uncertainty and measuring the likelihood of different events.

To make predictions using data analysis, one can use methods such as statistical mechanics, which was first introduced by Boltzmann. This method is based on the concept of entropy, which measures the amount of disorder or randomness as well as the information contained in a system. By applying statistical mechanics to a data set, one can determine the most likely outcomes and predict the future evolution of the system.

  What we propose to do is the following:  just take a data set  and make an analysis  with the goal to decompose  in subsets. Using the multiplicity of the different subsets to determine the probabilities for a realization of the different subsets. Then one may introduce some methods like statistical mechanics, which can be adopted multiscale entropic functional-the MED method. For example, if this  analysis of the data set  gives the following probability distribution  $Q(x^\prime,s^\prime|x,s)$  then we can use the multiscale entropic functional for a system,  which can be written in the following way (See a review on ED in \cite{e21100943} ): 
\begin{equation}
S[P,Q]=-\sum_{s^\prime}\int d^nx^\prime P(x^\prime,s^\prime|x,s)\log\frac{P(x^\prime,s^\prime|x,s)}{Q(x^\prime,s^\prime|x,s)}\,,\label{entropy-funA}
\end{equation}
Equation \eqref{entropy-funA} is the extension of the entropy functional in \cite{e21100943}.  As we are dealing with multiscale, the  additional ingredient is to sum over all scales. Here  $Q(x^\prime,s^\prime|x,s)$ is called the prior probability distribution,  which we get from our data set. The unknown function is the transition probability $P(x^\prime,s^\prime|x,s)$,  which  is analogous to  the transition probability distribution for a single particle  when it  moves from the position $x$ to a neighboring point $x^\prime$, where $s, s^\prime$ are scaling indices or arbitrary variables of the data set which is under consideration.  Our goal is to find the transition probability.     But first, we have to find the prior,  which we get from the analysis of the information data set,  which is under  the consideration.

The main idea of our approach is that in any system with time, the entropy rises, like in the II Law of thermodynamics,  which is dealing usually with a very complex system and a very large number of particles.    So we consider the dynamics of entropy having in mind this parallel with the Second Law of thermodynamics. Thus, our goal in this paper is to extend the application of ED\cite{e21100943} to a complex system { which can be described by some complex data set}. {Such a system can be also a single quantum mechanical particle embedded in different environments or scales}. The mathematical tools needed  for a system of { such directly noninteracting} particles go beyond the usual statistics and calculus. 
%\textcolor{red}{FK: many interacting particles required many-body correlated  function. It is a  special  very complex area of science,  which is called HNC and FHNC equations}.

 Because the system of free particles and any complex system involves various different constraints, and therefore one needs  to adopt a multifaceted and multi-scale approach to formulate the equations that describe the system accurately
 \cite{allen2017multiscale} (see, the references therein). 
It has applications in both natural and social sciences. The brain is a complex system of neurons. Society is a complex system. The universe itself is a complex system. It comprises galaxies. When zoomed in, a galaxy is a cluster of stars. When further zoomed in, there are solar systems, down to molecules and subatomic particles.%

In this paper, we consider a  simple   example of the extension entropic dynamic  to multiscale \cite{Caticha_2011, e21100943} with the goal to apply this method to Data Science, ie for the analysis of the data set and creating equations that may predict the evolutionary dynamics of the system. We call it multiscale entropic dynamics (MED) which can be applied to any dataset and may be used as new tool  for Data Science. Just for an illustration we are applying  MED to a simple single (non-interacting many) particle(s) systems, to obtain well-known Generalized Schr\"{o}dinger Equation (GSE). % for a system of particles.
%We develop this equation by taking inspiration from 
The obtained  equations  are identical with  numerous non-linear Schr\"{o}dinger equations (NLSE) that attracts the attention of the scientific community for more than  a century,  already. 
%equally of both mathematicians and physicists. 
%It is the case because, for mathematicians, NLSE is a system of coupled partial differential equations.
%Unsurprisingly, it poses a challenge to mathematicians to find solutions whether exactly or numerically. Physicists 
These equations describe non-linear phenomena in physics, e.g. as polarons, deformans, condensons,  optical or matter solitons and these fields and phenomena,  which are  still under intensive investigation in many new emerging fields,  systems and materials, e.g. as nonlinear optics, two-dimensional (2d) materials, cold and hot plasma physics, and crystal lattice dynamics. As one  example  the matter solitons are considered non-relativistic quanta of matter waves and represent Bose-Einstein Condensates (BEC) of atoms and electrons. Currently, researchers are taking a special interest in understanding the dynamics and formation of polarons,  self-trapped excitons, condenson, deformons, matter solitons in 2d materials, and many other nonlinear objects  \cite{Landau1933,Landau1933,zabusky1965interaction, fleischer2003observation,  zaera2018propagation, liu2020recent, feng2020mxene}.  Very possible that  first form of NLSE appeared in Landau Theory of phase transition where he first introduced  the cubic term in the Shroedinger  equation  describing order parameter\cite{Landau1937}. Later in 1950 this equation has been  applied by Ginzburg and Landau to superconductors and the NLSE got  a new meaning, the famous  Ginzburg-Landau equation\cite{ginzburg2009theory}.  Later in 1960s the idea  was applied to  BEC and the NLSE equation got the name as  Gross-Pitaevskii  equation. On the other hand an electron trapping by a crystal lattice,  which leads to other form of NLSE was also first described by Lev Landau in 1933 \cite{Landau1933}. Solomon Pekar proposed the concept of the polaron in 1946\cite{Pekar1946}, which was further developed by Landau and Pekar in a 1948 paper\cite{Pekar1948}. This theory suggested that polarons, not free electrons, were the charge carriers in ionic crystals. Unlike quantum electrodynamics, the polaron theory is free from divergences, and the electron energy and mass remain finite. Today, research on polarons continues to expand into new areas of 2D materials, where new forms of NLSE have been obtained\cite{sio2023polarons}. 
In general, it would be also interesting to consider these new phenomena from the principle of MED and compare them with conventional approaches.
%Hence there is an urgent need to develop a GSE that describes the dynamics of solitons in various scenarios. 

In particular, for a description of nonlinear waves  the NSLE was originally derived by Zabusky and Kruskal \cite{zabusky1965interaction}, but using MED one may include into the  Physics many non-equilibrium phenomena, dissipation, scattering and it can be used to describe the dynamics of solitons in 2d materials or many-body soliton physics. %However, the NLSE does not address some critical aspects of the dynamics of solitons such as soliton-soliton, and soliton-electromagnetic fields types of interactions in 2d materials. Therefore, a developed GSE must have the following salient features:
%1. GSE should have time dependence 
%2. GSE should be able to describe the dynamics of a system of decoupled and coupled solitons 
%3. GSE should be able to describe the interaction of solitons with electromagnetic fields. 

\section{The formulation of MED method and its application to Data Science %and GNSE
}\label{ED-sec1}
Our goal is to introduce a multiscale entropic dynamic method (MED) for Data Science.  For an illustration, we will apply MED to describe the motions of quantum particles. % in both linear and nonlinear regimes.
As an example of the complex  Data Science system  we  consider particles in %These fall into the category of
fractals. The fractals are self-similar structures that can be found in nature as well as constructed both experimentally and mathematically. Clouds, lightning, and coastlines are examples of natural fractals, and the Sierpinski triangle is an example of geometrical fractals \cite{doi:10.1063/1.2814878}. More geometrical objects that are fractals can be found in  Benoit B. Mandelbrot  foundational book on fractals \cite{MAN83}. The creation of solitons in Fractals has been discussed in  \cite{PhysRevE.61.R1048}.

%Here we present a set-theoretic formulation of fractals. We consider a fractal as a cluster that is a collection of sub-clusters, where each sub-cluster is a collection of sub sub-clusters, and so on. Mathematically, we have

For a quantum  statistical system, one must first specify the microstates, the prior probability distributions, and the constraints at the stage. For the  Data Science application, the prior probability distributions originated directly from the  existing data set  which is the subject of the main Data Science  analysis. The most important part  of this analysis is to find which constraints have been used  in the  collection of the existing data. The correct   evolution of entropy strongly depends on these  constraints. In the next step  we have to incorporate these limitations-ingredients in the entropy functional. With these taken into account we arrive at traditional  generalized  Boltzmann-like expression of entropy. % one obtains a desired formulation for depicting its %quantum mechanical 
representation of the  entropy. To derive the linear Schr\"{o}dinger equation (LSE), one considers $N$ noninteracting particles living in a flat Euclidean space. It is assumed that particles have definite initial positions (and  indefinite values of momenta) and yet unknown values that are desired to be inferred. The different definite initial positions of the particles are forming a data set.  Such a data set can be very large, that depends how many initial positions for a single particle we will take into account.
The  data set can be also split into subsets  associated with different scales, e.g. of the fractal.  %also have also be classified  for the 
%But here we deal with multiscale entropic dynamics (MED), where 
Note that  the  microstates at each scale are different. %We consider a supercluster which is a set of clusters given by 
%\begin{equation}
%G=\{G^s_1,G^s_2,\ldots,G^s_{N_s}\},\label{G1}
%\end{equation}
%where $s$ is the scaling index that represents the elements of $G$ and $N_s$ is the cardinality or the number of elements in $G$. The component $G^s_i$ has subcomponents
%\begin{equation}
%G^s_i=\{G^{s-1}_1,G^{s-1}_2,\ldots,G^{s-1}_{N_{s-%1}}\},\label{G2}
%\end{equation}
%The elements of this set also have subsubcomponents and so on.  A similar set notation was also used in \cite{e8030175}. 

%At each level, the elements of a given set are treated as point particles. This approximation is important in order to do physics. An example is the universe, where galaxies at a large scale are treated as point particles, at another scale the solar systems in galaxies are regarded as point particles, and so on. Our formalism is general that describes such a complex system.

Now we devise entropic dynamics (ED) at each scale. It is assumed that at each scale, the particle lives in a Euclidean space $\mathcal{X}_s$ with metric $\delta_{ab}$, with $a=1,2,3$ for spatial coordinates. And for all particles at that scale we have $\mathcal{X}_{N_s}=\mathcal{X}_s\times\ldots\mathcal{X}_s$, which is $3N_s$ dimensional configuration space.  The positions of the particles are given by $x^a_i\in\mathcal{X}_{N_s}$, where the index $i=1,2,\ldots N_s$. We represent $x^a_i$ collectively by $x$.

The multi-scale entropic functional for a system can be written in the following way ( See a review on ED in \cite{e21100943}): 
\begin{equation}
S[P,Q]=-\sum_{s^\prime}\int d^nx^\prime P(x^\prime,s^\prime|x,s)\log\frac{P(x^\prime,s^\prime|x,s)}{Q(x^\prime,s^\prime|x,s)}\,,\label{entropy-fun}
\end{equation}
Equation \eqref{entropy-fun} is the extension of the entropy functional in \cite{e21100943}.  As we are dealing with multi-scale, the  additional ingredient is to sum over all scales. Here $Q(x^\prime,s^\prime|x,s)$ is the prior probability distribution, and $P(x^\prime,s^\prime|x,s)$ is the transition probability distribution as the particle moves from $x$ to a neighboring point $x^\prime$, where $s, s^\prime$ are scaling indices.  Our goal is to find the transition probability.     But first, we have to determine the prior probability  distribution. For  any specific data-set, it can be obtained directly  by classifying different snapshots of the data-set. For the particular case of our many-particle system  as for the ideal gas to determine prior we will follow the original Boltzmann approach.
%
%and wish to find the transition probability 
  %To find $P$, we maximize the entropic functional %subject to various constraints as follows.
%
%\subsection{The Prior}
The prior probability distribution $Q(x^\prime,s^\prime|x,s)$ codifies the relation between $x$ and $x^\prime$ before the information contained in constraints has been processed,  where all particle positions are equally probable. In other words, it is desired to find a prior distribution that is invariant under translation and rotation. It can be obtained by maximizing the following relative entropy.
 
\begin{equation}
S(Q)=-\sum_{s^\prime}\int d^nx^\prime Q(\Delta  x)\log\frac{Q(\Delta x)}{\mu(\Delta x)}\,,
\end{equation}
where $\Delta x=x^\prime -x$ is relative to the uniform measure $\mu(\Delta x)$, subject to normalization and a constraint that concerns short steps,
\begin{equation}
\sum_{s^\prime}\int d^nx^\prime Q(x^\prime,s^\prime|x,s)\delta_{ab}\Delta x_i^a\Delta x_i^b=\langle \Delta \ell^2_i \rangle\,, \ (i=1,2,\ldots N_{s^\prime})
\end{equation}
where $\langle\Delta \ell^2_i \rangle$ are constants equal to the square of the average displacement between the points $x$ and $x^\prime$. The index $i$ indicates that there are $N_{s^\prime}$ constraints at each scale and which are rotational invariant.   

The result is 
\begin{equation}
Q(x^\prime,s^\prime|x,s)\propto\exp[-\frac{1}{2}\mathop{\textstyle\sum}_{s^\prime}\mathop{\textstyle\sum}_{i}^{N_{s^\prime}}\frac{1}{\sigma^2_{s^\prime,i}}\delta_{ab}\Delta x^a_i\Delta x^b_i]\,,\label{prior1}
\end{equation}
where $\sigma^2_{s^\prime,i}$  is a Lagrange multiplier which will be determined later (see equation \eqref{sigma1}.To ensure small steps, this Lagrange multiplier must be very small. The right-hand side is the product of the Gaussian function which means the short steps are independent of each other.    Equation \eqref{prior1} is a prior probability distribution that only takes into account the original positions of a particle before the actual constraints or information is incorporated such as the influence of the EM field. It only describes motion in short steps as the particle moves from $x$ to $x^\prime$. 

\subsection{Application of Multi-scale Entropy Dynamic to a Quantum Mechanical System of Particles }\label{applicatio-MED}
We want to write down a form of equation \eqref{entropy-fun} that works for describing the dynamics of particles and quasi-particles in solids considering the case when the  isotropic symmetry of the space is broken, e.g. by applying an external electric field. This symmetry breaking requires including extra constraints that are listed below. In this way, the transition probability distribution $P(x^\prime,s^\prime|x,s)$ will be determined.

\begin{enumerate}
\item In order to introduce the breaking of the isotropic symmetry of the space the  external electric field  
%and also to account for coupling, we impose that the expected displacement in the direction of 
is presented  by the gradient of a certain ”potential” $\phi_s(x^a)$,  which satisfies to the constraint:  

\begin{equation}
\sum_{s^\prime}\int d^nx^\prime P(x^\prime,s^\prime|x,s)\Delta x^a\frac{\partial\phi_s}{\partial x^a}=\kappa_{1,s}\,. \label{phi-pot}
\end{equation}
This constraint is called the drift potential constraint. Here $\kappa_{1,s}$ are constants which are related to equipotential energy at the Fermi Surface of a Solid. Without an applied electric field the Fermi surface is the equipotential energy of the (quasi-)particles. But with the application of an electric field, the equipotential energy is related to the cross-section of the Fermi surface,  which is perpendicular to the applied field.
\item The interaction of a charged particle with an external electromagnetic (EM) field  is represented by imposing the constraint 
\begin{equation}
\sum_{s^\prime}\int d^nx^\prime P(x^\prime,s^\prime|x,s)\Delta x^aA_a=\kappa^s_2\,,\label{vec-pot}
\end{equation}

where $A_a$ is vector potential and $\kappa^s_2$ are constant and it represents average displacement in the direction of the vector potential. It should be noted that the value of vector potential has usually no meaning unless it can be changed by a gauge transformation. The vector potential further constrains the trajectory of the particle. This constraint is called gauge constraint. Although it is not gauge invariant, but it is possible to fix the gauge in the way done in\cite{caticha2019entropic}. We leave the gauge issue here as it is, and will proceed to derive the transition probability without it, see equation \eqref{Prob2}
Nevertheless, the invariant form of gauge constraint may be written in the following form:
\begin{equation}
\sum_{s^\prime}\int d^nx^\prime P(x^\prime,s^\prime|x,s)\Delta x_a\epsilon^{abc}  \frac{\partial A_c}{\partial x^b}=\kappa_{1,s}=\kappa^s_2\,,\label{vec-pot2}
\end{equation}
   
\end{enumerate}
The maximized entropy functional of equation \eqref{entropy-fun} subject to the constraints are given in equations \eqref{phi-pot}, \eqref{vec-pot}. We get the following result for the transition probability by combining both constraints. 
\begin{equation}
P(x^\prime,s^\prime|x,s)\propto\exp\left[-\frac{1}{2}\mathop{\textstyle\sum}_{s^\prime}\mathop{\textstyle\sum}_{i}^{N_{s^\prime}}\left(\frac{1}{\sigma^2_{s,i}}\delta_{ab}\Delta x_i^a\Delta x_i^b-\alpha_{s,i}^\prime\Delta x_i^a\frac{\partial\phi_s}{\partial x_i^a}+\beta_{s^\prime,i}\Delta x_i^a A_a\right)\right] 
\,,\label{Prob2}
\end{equation}
where $\sigma^2_{s,i}$, $\alpha^\prime_{s,i}$, and $\beta_{s^\prime,i}$  
are Lagrange multipliers which will be expressed in the form of Planck's constant $\hbar$, speed of light, and charge of an electron. 
For the sake of completeness, we should note that the gauge 
invariance of equation \eqref{Prob2} can be achieved %gauge 
and may be written in the following way:
\begin{equation}
P(x^\prime,s^\prime|x,s)\propto\exp\left[-\frac{1}{2}\mathop{\textstyle\sum}_{s^\prime}\mathop{\textstyle\sum}_{i}^{N_{s^\prime}}\left(\frac{1}{\sigma^2_{s,i}}\delta_{ab}\Delta x_i^a\Delta x_i^b-\alpha_{s,i}^\prime\Delta x_i^a\frac{\partial\phi_s}{\partial x_i^a}+\beta_{s^\prime,i}\Delta x_a\epsilon^{abc}  \frac{\partial A_c}{\partial x^b}=\kappa_{1,s}\right)\right] 
\,,\label{Prob1}
\end{equation}
%}

In the MED formulation, we derive below the transition probability in the Gaussian form with time evolution. In this formulation, Brownian motion was applied explicitly to the general form transition probability as well as  entropic time. Entropic time is explained and derived next. 

Any notion of time must involve motion and change\cite{caticha2001change}. In MED, motion/change is described by the transition probability given by \eqref{Prob2}. It is desired to obtain small change. Large changes can be obtained by accumulating small short steps. It should be noted that any notion of time must have  (a) something one might identify as an instant, (b) a sense in which these instants can be ordered, (c) a convenient concept of duration measuring the separation between instants \cite{caticha2011entropic}. In ED an instant is defined by the information required to generate the next instant. The point $x$ occurs at time $t$ and $x^\prime$ occurs at $t^\prime$. So the probability distribution evolves according to
\begin{equation}
\rho(x^\prime,s^\prime,t^\prime)=\sum_s\int dt dx P(x^\prime,s^\prime|x,s)\rho(x,s,t).
\end{equation}
We write $\rho(x,s,t)=\rho_s(x,t)$. Having introduced the notion of time, the next step is to define a duration of time. Since our goal is to derive GSE, it suffices to construct a Newtonian interval that is independent of the position $x$ and time $t$. This can be achieved by the Lagrange multiplier $\sigma^2_{s, i}$ to be constant such that
\begin{equation}
\frac{1}{\sigma^2_{s, i}}=\frac{m_{i,s}}{\eta_s\Delta t} \label{sigma1}
\end{equation}
where $m_{i,s}$ are the particle masses and $\eta_s$ is a constant which will be shown later that it is $\hbar$. Furthermore
\begin{equation}
M_{ab}=m_{s,i}\delta_{ab}\,,\label{mass-matrix}
\end{equation}
where $M_{ab}$ is effective mass matrix. We have
\begin{equation}
P(x^\prime,s^\prime|x,s)\propto\exp\left[-\frac{1}{2\eta_s\Delta t}M_{ab}(\Delta x^a-\langle\Delta x^a\rangle)(\Delta x^b-\langle\Delta x^b\rangle)\right]
\,.
\end{equation}
Here
\begin{equation}
\Delta x^a =\langle \Delta x^a\rangle +\Delta w^a\,,
\end{equation}
with
\begin{equation}
\langle \Delta x^a\rangle=\eta_s\Delta t M^{ab}\left(\alpha^\prime_{s,i}\frac{\partial\phi_s}{\partial x^b}-\beta_sA_b\right)\,,
\end{equation}

%Delta x_a\epsilon^{abc}  \frac{\partial A_c}{\partial x^b}

\begin{equation}
\langle\Delta w^a\rangle=0\, \ \ \text{and}\ \  \langle\Delta w^a\Delta w^b\rangle=\eta_s\Delta t M^{ab}\,. \label{Prob}
\end{equation}
This is Brownian motion because the drift $\langle \Delta x^a\rangle\sim O(\Delta t)$ and the fluctuation $\Delta w^a\sim O(\Delta t^{1/2})$. The trajectory is continuous but not differentiable.

The probability $\rho_s(x,t)$ evolves according to the Fokker-Planck (FP) equation. 
\begin{eqnarray}
\frac{\partial\rho_s}{\partial t}=-\partial_a(v^a_s\rho_s)
\,.\label{FP1}
\end{eqnarray} 
Note that $s$ is the scaling index and $a=1,2.3$ are spatial indices. A summation over repeated indices should be understood. Here $v^a$ is the current velocity given by 
\begin{equation}
v^a_s=M^{ab}(\alpha^\prime_{s,i}\partial_a\Phi_s-\beta_sA_a)\,, \text{where}\ \ \Phi_s=\alpha^\prime_{s,i}\eta_s\phi_s-\eta_s\log\rho_s^{1/2}
\end{equation}
 %Note that the current velocity is invariant under the gauge transformation $\Phi\to\Phi+\beta_{s^\prime,s}\chi$ and $A_a\to A_a+\alpha^\prime_{s^\prime,s}\partial/\partial x^a_{i}\chi$
%\section{Special Cases of GSE}
%\subsection{}

So far we only have one dynamical variable $\rho_s$ which evolves according to Fokker-Planck (FP). To derive the Schr\"{o}dinger equation we need two dynamical variables the probability $\rho_s$ and phase $\Phi_s$. To promote $\Phi_s$ to a dynamical variable, we need another constraint $H=H(\rho_s,\Phi_s)$, where $H$ is an energy functional. By requiring that the energy is conserved, we obtain the second dynamical variable as well. The functional $H(\rho_s, \Phi_s)$ can be constructed by writing the FP equation as
\begin{equation}
\frac{\partial\rho_s}{\partial t}=\frac{\delta H}{\delta \Phi_s}\,.\label{FP2}
\end{equation}
It can easily be checked that the appropriate energy function is given by
\begin{equation}
H(\rho_s,\Phi_s)=\sum_s\int dx \rho_s\left(\frac{1}{2}M^{ab}(\partial_a\Phi_s-\beta_sA_a)(\partial_b\Phi_s-\beta_sA_b)+V_s(x)\right)+\mathop{\textstyle\sum}_{ss^\prime}g_{ss^\prime}F(\rho_s,\rho_{s^\prime})\,,\label{H1}
\end{equation}
where $V(x)$ is a scalar potential and  $F(\rho_s,\rho_{s^\prime})$ is an integration to be determined below and $g_{ss^\prime}$ is the complexity coefficient. This term leads to the nonlinear Schr\"{o}dinger equation.  We have
\begin{equation}
\frac{\delta H}{\delta \rho_s}=\frac{1}{2}M^{ab}(\partial_a\Phi_s-\beta_sA_a)(\partial_b\Phi_s-\beta_sA_b)+V_s(x)+
\mathop{\textstyle\sum}_{s^\prime}g_{ss^\prime}\frac{\delta F(\rho_s,\rho_{s^\prime})}{\delta\rho_s}
\end{equation}
Taking total time derivative of Eq. \eqref{H1} and require it to be conserved  and also incorporate Eq.\eqref{FP2}
\begin{equation}
\frac{dH}{dt}=\sum_s\int dx\left[\frac{\delta H}{\delta\Phi_s}\partial_t\Phi_s+\frac{\delta H}{\delta \rho_s}\partial_t\rho_s\right]=\sum_s\int dx\left[\partial_t\Phi_s+\frac{\delta H}{\delta \rho_s}\right]\partial_t\rho_s=0
\end{equation}
It holds for all $\partial_t\rho_s$ which means that
\begin{equation}
\frac{\partial\Phi_s}{\partial t}=-\frac{\delta H}{\partial\rho_s}\,.\label{HJ1}
\end{equation}
We get
\begin{equation}
\frac{\partial\Phi_s}{\partial t}=-\frac{1}{2}M^{ab}(\partial_a\Phi_s-\beta_sA_a)(\partial_b\Phi_s-\beta_sA_b)-V_s-\mathop{\textstyle\sum}_{s^\prime}g_{ss^\prime}\frac{\delta F(\rho_s,\rho_{s^\prime})}{\delta \rho_s}\,.\label{HJ2}
\end{equation}
This is the quantum Hamilton-Jacobi equation.  Eqs. \eqref{FP1} and \eqref{HJ2} can be combined using
\begin{equation}
\psi_s=\rho_s^{1/2}\exp[ik\Phi_s/\eta_s]\,.\label{psi}
\end{equation}
The result is
\begin{eqnarray}
\frac{i\eta_s}{k}\frac{\partial\psi_s}{\partial t}&=&\frac{\eta^2_s}{2k^2}M^{ab}(i\partial_a-\beta_sA_a)(i\partial_b-\beta_sA_a)\psi_s+V_s\psi_s
+\frac{\eta_s^2}{2k^2}\frac{M^{ab}\partial_a\partial_b\sqrt{\rho_s}}{\sqrt{\rho_s}}\psi_s\notag\\
&+&\mathop{\textstyle\sum}_{s^\prime} g_{ss^\prime}\frac{\delta F(\rho_s,\rho_{s^\prime})}{\delta\rho_s}\psi_s\,,\label{pre-gauge}
\end{eqnarray}

The physical meaning of the $\psi_s$ in equation \eqref{psi} is that it represents the wavefunction of the particle that generalized Schr\"{o}dinger equation. Its modulus is the probability of finding the particle in space and time. In most general situations e.g. data science, the 
$\psi_s$ will be the parameter controlling the transition probabilities.  
Note that equation \eqref{pre-gauge} is invariant under the gauge transformation given below \cite{caticha2021entropic} 
\begin{equation}
\psi_s\to\psi^\prime_s=e^{i\beta\chi(x,t)}\psi_s\ \ \text{and}\ \ A_a\to A_a^\prime=A_a+\partial_a\chi.  
\end{equation}
In equation \eqref{pre-gauge}, the third term on the right is called `quantum potential'. Normally this term is present in the Hamilton-Jacobi equation \eqref{HJ2}. On combining this equation with the Fokker-Planck equation  \eqref{FP1}, one obtains linear Schr\"{o}dinger equation (LSE) that obeys the superposition principle. In our case, the quantum potential is implicit in $F(\rho_s,\rho_{s^\prime})$. Since we have freedom in the choice of  $F$, we choose it such that
\begin{equation}
\mathop{\textstyle\sum}_{s^\prime} g_{ss^\prime}\frac{\delta F(\rho_s,\rho_{s^\prime})}{\delta\rho_s} +\frac{\eta_s^2}{2k^2}\frac{M^{ab}\partial_a\partial_b\sqrt{\rho_s}}{\sqrt{\rho_s}}=\mathop{\textstyle\sum}_{s^\prime} g_{ss^\prime}f(\rho_{s^\prime})\,.\label{F-0}
\end{equation}
Note the function $f$ on the right as a function of one variable only. If the goal is to obtain an LSE, one can set $f=0$. However, we are interested in nonzero $f$ for the reasons given below.  
\begin{equation}
i\hbar\frac{\partial\psi_s}{\partial t}=\frac{\hbar^2}{2}M^{ab}(i\partial_a-\frac{e}{\hbar c}A_a)(i\partial_b-\frac{e}{\hbar c}A_a)\psi_s+V_s\psi_s
+\mathop{\textstyle\sum}_{s^\prime} g_{ss^\prime}f(\rho_{s^\prime})\psi_s\,.\label{SEa1}
\end{equation}
Here we used $\eta_s/k=\hbar$, and $\beta_s=e/\hbar c$, where $e$ is the charge of an electron $c$ is the speed of light.  Equation \eqref{SEa1} is the sought GSE that takes into account the EM field interaction with matter waves. The last term indicates nonlinearity. A similar last term is also reported in \cite{PhysRevE.61.R1048}. But here we naturally derived the general NLSE using Entropic dynamics. For solitons we can take $f(\rho_{s^\prime})=\rho_{s^\prime}=|\psi_{s^\prime}|^2$. So that 
\begin{equation}
i\hbar\frac{\partial\psi_s}{\partial t}=\frac{\hbar^2}{2}M^{ab}(i\partial_a-\frac{e}{\hbar c}A_a)(i\partial_b-\frac{e}{\hbar c}A_b)\psi_s+V_s\psi_s
+\mathop{\textstyle\sum}_{s^\prime} g_{ss^\prime}|\psi_{s^\prime}|^2\psi_s\,.\label{SE1}
\end{equation}
The existence of the nonlinear terms\cite{gulevich2006new},\cite{gulevich2006perturbation},\cite{gulevich2008shape}  and, in particular, cubic term\cite{kusmartsev1984classification} is characteristic of the solitons and other nonlinear waves (see, the seminal paper by Zakharov\cite{zakharov1968stability} about nonlinear stability of periodic waves in deep water). The existence of such waves and stable solitons depends on  the boundary conditions and their  dynamical stability the depends on the spatial dimension of the system\cite{zakharov1974three},\cite{kusmartsev1984classification}. Besides solitons there is a large variety of nonlinear phenomena, including shape waves \cite{gulevich2008shape}\cite{gulevich2009shape}, periodic waves in deep and shallow water\cite{zakharov1968stability}, plasma cavitons\cite{zakharov1974three}\cite{kusmartsev1983self},  Urbach and Lifshiz density of states tails\cite{kusmartsev1991multiphonon}\cite{kusmartsev1984disappearing}, the collapse of the plasmon, Langmuir waves\cite{zakharov1972collapse} and many others related phenomena\cite{kusmartsev1984symmetry} \cite{fomin1988non}. It is very interesting if  the NSE equations describing these or associated phenomena can be obtained with the principle of the   maximum entropy and entropic dynamics as described above.

\section{Reduction of GNSE to Some Relevant Forms}\label{solitons-1}
One can note that equation \eqref{SE1} is a complex system of equations.  It has some general form,  which covers numerous physical phenomena. %s although some components are still missing.
It may have both scalar or vector(tensor) forms\cite{zakharov1972collapse} \cite{kusmartsev1989application}. Below we will  discuss those forms of NLS which found direct applications in different areas of physics,  which covers  not only  solitons\cite{novikov1984theory}\cite{kusmartsev1984classification}\cite{kusmartsev1989application} but other phenomena such as  self-trapping and polarons\cite{fomin1988non}\cite{kusmartsev1984self}\cite{kusmartsev1981jahn} as well as plasma caviton formation\cite{zakharov1985hamiltonian}\cite{kusmartsev1983self}\cite{kusmartsev1983zh}.
Here we consider the simplest example where two solitons coupled to each other may be created\cite{scott1992davydov}. In one case, we obtain the vector nonlinear Schr\"{o}dinger equation (VNSE)\cite{kusmartsev1981jahn}. In the other case, the scalar nonlinear Schr\"{o}dinger equation (SNSE) is obtained. The difference between VNSE and SNSE is that the former involves coupled solitons and the latter involves decoupled solitons as well as many different physical phenomena.

\subsection{Scalar Form of GSE for Decoupled Solitons and Electron and Exciton Self-trapping }

The solitons  do usually exist  in one-dimensional chain or system with reduced dimensions\cite{kusmartsev1984classification}. The illustrative example is Davydov solitons\cite{scott1992davydov}, created in protein chains. They are associated with electron self-trapping or localization of Amide-I (or CO stretching) vibrational energy in proteins. Such localization as well as electron self-trapping arises through the interaction of the Amide-I mode with lattice distortion and plays an important role in charge transport vital for all biological systems. 
Our starting point is the system \eqref{SE1}. For illustration, the EM field is set to zero ($\vec A=0$). The SNSE can be obtained by setting 
\begin{equation}
g_{ss^\prime}=0, \ \ \text{when}\ \ s\not=s^\prime
\end{equation}
where  $g_{11}$ and $g_{22}$ survive. Further set $\eta_s/k=\hbar$. We obtain two decoupled SNSE's as follows
\begin{equation}
i\hbar\frac{\partial \psi_1}{\partial t}=-\frac{\hbar^2}{2m}\nabla^2\psi_1+V_1\psi_1+ g_{11}|\psi_1|^2\psi_1\,.\label{SNSE-1}
\end{equation}
\begin{equation}
i\hbar\frac{\partial \psi_2}{\partial t}=-\frac{\hbar^2}{2m}\nabla^2_i\psi_2+V_2\psi_2+ g_{22}|\psi_2|^2\psi_2\,,\label{SNSE-2}
\end{equation}
which is the desired system of two decoupled solitons. The last two equations are known as the Gross-Pitaevski equation (GPE) \cite{soljacic2000self, damgaard2021scattering}. In Bose-Einstein condensate (BSE) $g<0$ is referred to as the bright solitons and $g>0$ is called dark solitons \cite{damgaard2021scattering}.

\subsection{Vector Form of GSE for Coupled Solitons}
We again recall the system \eqref{SE1} with $\vec A=0$, for illustration. Set $g_{ss^\prime}$ such that
\begin{equation}
g_{ss^\prime}=0, \ \  \text{when}\ \ s=s^\prime\,,
\end{equation}

where $s, s^\prime=1,2$.  Equation  \eqref{SE1} simplifies to 
\begin{equation}
i\hbar\frac{\partial \psi_1}{\partial t}=-\frac{\hbar^2}{2m}\nabla^2\psi_1+V_1\psi_1+ g_{12}|\psi_2|^2\psi_1\,.\label{VNSE-1}
\end{equation}
\begin{equation}
i\hbar\frac{\partial \psi_2}{\partial t}=-\frac{\hbar^2}{2m}\nabla^2\psi_2+V_2\psi_2+ g_{21}|\psi_1|^2\psi_2\,,\label{VNSE-2}
\end{equation}
which is the desired system of two solitons. Generally, we have
 
\begin{equation}
i\hbar\frac{\partial \psi_1}{\partial t}=-\frac{\hbar^2}{2m}\nabla^2\psi_1+V_1\psi_1+\hbar g_{11}|\psi_1|^2\psi_1+ g_{12}|\psi_2|^2\psi_1\,.\label{VNSE-1ab}
\end{equation}
\begin{equation}
i\hbar\frac{\partial \psi_2}{\partial t}=-\frac{\hbar^2}{2m}\nabla^2\psi_2+V_2\psi_2+\hbar g_{21}|\psi_1|^2\psi_2+g_{22}|\psi_2|^2\psi_2\,,\label{VNSE-2cd}
\end{equation}

\subsection{Interaction of Electromagnetic Fields with Coupled and Decoupled Solitons}
We can also write the full equation for solitons with EM-field. Recall equation \eqref{SE1}
\begin{equation}
i\hbar\frac{\partial\psi_s}{\partial t}=\frac{\hbar^2}{2}M^{ab}(i\partial_a-\frac{e}{\hbar c}A_a)(i\partial_b-\frac{e}{\hbar c}A_a)\psi_s+V_s\psi_s
+\mathop{\textstyle\sum}_{s^\prime} g_{ss^\prime}|\psi_{s^\prime}|^2\psi_s\,.\label{SE-EM-1}
\end{equation}
where $M^{ab}=\delta^{ab}/m_i$ is the the inverse of mass matrix. For coupled solitons we have
\begin{equation}
i\hbar\frac{\partial\psi_1}{\partial t}=\frac{\hbar^2}{2m}\delta^{ab}(i\partial_a-\frac{e}{\hbar c}A_a)(i\partial_b-\frac{e}{\hbar c}A_a)\psi_1+V_1\psi_1
+ g_{12}|\psi_{2}|^2\psi_1\,.\label{coupled-1}
\end{equation}
\begin{equation}
i\hbar\frac{\partial\psi_2}{\partial t}=\frac{\hbar^2}{2m}\delta^{ab}(i\partial_a-\frac{e}{\hbar c}A_a)(i\partial_b-\frac{e}{\hbar c}A_a)\psi_2+V_2\psi_2
+ g_{21}|\psi_{1}|^2\psi_2\,.\label{coupled-2}
\end{equation}

Similarly, for decoupled solitons we have
\begin{equation}
i\hbar\frac{\partial\psi_1}{\partial t}=\frac{\hbar^2}{2m}\delta^{ab}(i\partial_a-\frac{e}{\hbar c}A_a)(i\partial_b-\frac{e}{\hbar c}A_a)\psi_1+V_1\psi_1
+ g_{11}|\psi_{1}|^2\psi_1\,.\label{decoupled-1}
\end{equation}
\begin{equation}
i\hbar\frac{\partial\psi_2}{\partial t}=\frac{\hbar^2}{2m}\delta^{ab}(i\partial_a-\frac{e}{\hbar c}A_a)(i\partial_b-\frac{e}{\hbar c}A_a)\psi_2+V_2\psi_2
+ g_{22}|\psi_{2}|^2\psi_2\,.\label{decoupled-2}
\end{equation}
In summary, equations \eqref{SNSE-1} through \eqref{decoupled-2} are special cases of equation \eqref{SE1}. Generally, the scaling indices $s, s^\prime$ may vary as $1,2,\ldots n$ that describes a system of $n$ coupled equations or quasi-particles. It is also worthwhile to note that the coupling enters through the complex coefficient $g_{ss^\prime}$. If it is set to zero, the equation \eqref{SE1} reduces to the linear SE that obeys the usual superposition principle for $n$ particles system.

\section{Discussion}\label{Discussion}
In the 1950s, Ginsburg and Landau proposed a functional called the Ginsburg-Landau (GL) functional for free energy to describe the superconducting state in solids \cite{ginzburg2009theory}. The GL functional was developed to introduce a non-linear term in the Schr\"{o}dinger equation (SE) to describe the superconductivity property of conductors, by presenting the conduction electrons as super-fluids. The minimization of this GL functional gives rise to the nonlinear Schr\"{o}dinger equation (NLSE).

The NLSE describes a new state of quasi-particles, the superconducting condensate, which is similar to Bose-Einstein condensation (BEC) \cite{zabusky1965interaction}. Gross and Pitaevskii later derived the NLSE by applying the minimum energy principle to the free energy of electrons or atoms in BEC. The use of NLSE enabled the description of the quantum dynamics of other systems in the form of solitary matter waves or solitons in coupled solitons or containing multi-solitons, or having optically interacting solitons.

The NLSE has been widely used in many fields of physics, including condensed matter physics, nonlinear optics, and fluid dynamics. Solitons, which are localized wave packets that maintain their shape during propagation, have been described by the NLSE. These solitons can exist in many different systems, such as in optical fibers, plasmas, and superfluids. 

The maximization of MED functional not only resulted in an extension of GPE, but it provides a natural way to include other interactions in it, such as the interaction of electromagnetic fields with quasi-particles in solids. Further, it provides the tools to deal with the dynamics of the scalar as well as vector solitons in decoupled and coupled forms. Whereas, it is to be noted that the GPE deals with scalar solitons. The implications of the GNSE may be quite far-reaching in our opinion. Its application to 2d materials may lead to opportunities for discovering the quantized energies of solitons at the point defects sites of those materials. Such quantized states may turn out to be suitable for future applications in electronics including quantum computing.

\section{Conclusions}\label{conclusions}
The MED formulation was successfully applied to the derivation of GNSE which represents non-relativistic quantum mechanics in both linear and non-linear forms. The latter form describes the dynamics of quasi-particles. As an example, it describes the dynamics of matter-solitons in 2d materials for coupled and decoupled solitons.

Furthermore, the development of machine learning and artificial intelligence techniques has further strengthened the connection between these fields. These techniques, which are used extensively in data science, are based on statistical and probabilistic models
 that are similar to those used in statistical physics and information theory. For example, deep learning models are based on neural networks that are similar to those used in statistical physics to model the behavior of physical systems.

In a recent paper by analysis of the huge database, the relationship between DNA methylation and mutability, specifically how methylation can affect the emergence of novel genetic variations in eukaryotes (organisms with cells containing a nucleus) has been identified\cite{kusmartsev2020cytosine}. Further analysis of somatic mutation data, particularly from cancers where specific repair pathways are compromised, is necessary to understand the underlying mechanisms of this process and the involvement of particular DNA repair pathways as well as how  the impact of methylation on mutability extends beyond the methylated cytosine itself. DNA methylation, which is a common epigenetic modification in eukaryotes, can affect genetic variation in ways that are not fully understood and imply that the precise mechanisms involved are complex and require further investigation\cite{kusmartsev2020cytosine}. Their findings suggest that methylation has a significant impact on the emergence of novel genetic variants in eukaryotes, which may have important implications for understanding genetic diversity and disease. We hope that the proposed EMD  method to the huge databases can further shed the light on its evolutional mechanisms\cite{kusmartsev2020cytosine}.

Additionally, all three fields are concerned with the extraction of meaningful information from complex and noisy data sets. Statistical physics seeks to identify and understand the underlying patterns and structures that govern the behavior of physical systems.
 Information theory seeks to extract and transmit useful information from noisy or uncertain data sets. Data science seeks to extract insights and knowledge from large and complex data sets.

In summary, the fields of statistical physics, information theory, and data science are connected through their use of similar mathematical and statistical tools to model and analyze complex systems. These connections have become even stronger with the development
 of machine learning and artificial intelligence techniques, which rely heavily on the probabilistic and statistical models developed in these fields. The MED method can be adopted to improve the accuracy of data analysis and involve in optimizing the subset selection process to minimize the error in the prediction of the system's evolution. We expect that this method will be effective in a wide range of applications, including finance, healthcare, and social media analysis

\section*{Acknowledgement}
%\textbf{Acknowledgment}\\
SN would like to thank A.~Caticha for many insightful discussions on entropic dynamics and especially for his comment on the role of momentum in ED. This work is partially supported with funds received under ADEK grant AARE-2019-131, and with the funds received under Khalifa University's grant ESIG-2023-003. FVK  is grateful to 1000 Talents Award, China;  PIFI CAS fellowship;  Khalifa University of Science and Technology under Award Nos FSU-2021-030/8474000371 and the EU H2020 RISE project TERASSE (H2020-823878).

%\bibliography{ref3}
%

\end{document}